%% file: main.tex
\documentclass[conference]{IEEEtran}
\IEEEoverridecommandlockouts
\usepackage{amsmath}
\usepackage{pxfonts}
\usepackage{graphicx}
\usepackage{cite}
\usepackage{algorithm2e}
\usepackage{epsfig}

\newtheorem{lemma}{Lemma}

\begin{document}
%
\title{Collaborative P2P Streaming of \\
Interactive Live Free Viewpoint Video
}


\author{
\IEEEauthorblockN	{\itshape 
	Dongni Ren$^\ast$   \qquad
	S.-H. Gary Chan$^\ast$  \qquad
	Gene Cheung$^{\dagger}$ \qquad
	Vicky Zhao$^{\ddagger}$ \qquad
	Pascal Frossard$^{\star}$ \qquad
	}
\IEEEauthorblockA{
	$^\ast$The Hong Kong University of Science and Technology, 
        Hong Kong, China\\
	$^{\dagger}$National Institute of Informatics, 
	Tokyo, Japan \\
	$^{\ddagger}$University of Alberta, Edmonton, AB, Canada\\
	$^{\star}$\'Ecole Polytechnique F\'ed\'erale de Lausanne (EPFL), 
	Lausanne, Switzerland\\
	Email: \{tonyren, gchan\} @cse.ust.hk$^\ast$, cheung@nii.ac.jp$^{\dagger}$, vzhao@ece.ualberta.ca$^{\ddagger}$, pascal.frossard@epfl.ch$^{\star}$\\
}
}

%


\maketitle

\begin{abstract}
We study an interactive live streaming scenario where multiple
peers pull streams of the same free viewpoint video that are 
synchronized in time but not necessarily in view. In free
viewpoint video, each user can periodically select a virtual view
between two anchor camera views for display. The virtual view is
synthesized using texture and depth videos of the anchor views via
depth-image-based rendering (DIBR).  In general, the distortion of the
virtual view increases with the distance to the anchor views, and
hence it is beneficial for a peer to select the
closest anchor views for synthesis.  On the other hand, if peers
interested in different virtual views are willing to tolerate larger
distortion in using more distant anchor views, they can collectively
share the access cost of common anchor views.

Given anchor view access cost and synthesized distortion of virtual
views between anchor views, we study the optimization of anchor view 
allocation for collaborative peers. We
first show that, if the network reconfiguration costs due to
view-switching are negligible, the
problem can be optimally and efficiently solved in polynomial time
using dynamic programming.  We then consider the case of 
non-negligible reconfiguration costs 
({\it e.g.}, large or
frequent view-switching
leading to anchor-view changes).
In this case, the view allocation problem becomes NP-hard. We thus
present a locally optimal and centralized allocation algorithm
inspired by Lloyd's algorithm in non-uniform scalar quantization. We
also propose a distributed algorithm with guaranteed 
convergence where each
peer group independently make merge-and-split decisions with a
well-defined fairness criteria.  The results show that depending on the
problem settings, our proposed algorithms achieve respective optimal and 
close-to-optimal performance in terms
of total cost, and substantially outperform a P2P scheme without
collaborative anchor selection.

\end{abstract}


%
\IEEEpeerreviewmaketitle

\section{Introduction}
\label{sec:intro}
\input{intro}

\section{Related Work}
\label{sec:related}
\input{related}

\section{Collaborative Streaming Model}
\label{sec:system}
\input{system}

\section{Formulation I: no reconfiguration cost}
\label{sec:formulate}

\input{formulate}

\subsection{Algorithm I: DP solution}
\label{sec:algo}
\input{algorithm}

\section{Formulation II: reconfiguration cost}
\label{sec:formulate2}
\input{formulate2}

\section{Algorithm II: heuristics}
\label{sec:algo2}
\input{algo2}

\section{Experimentation}
\label{sec:results}
\input{results}

\section{Conclusion}
\label{sec:conclude}
\input{conclude}



\bibliographystyle{IEEEtran}
\bibliography{ref}

\begin{small}
\appendix
\input{append1}

\end{small}

\end{document}

%% file: intro.tex
The advent of multiview imaging technologies means that videos from
different viewpoints of the same 3D scene can now be captured 
simultaneously by a system of multiple closely spaced 
cameras~\cite{fujii06}. 
If depth maps (per-pixel distance between camera and physical objects) 
from the same camera viewpoints are also 
available,\footnote{Depth maps can be captured directly through 
time-of-flight (ToF) cameras~\cite{gokturk04}, or indrectly through 
stereo-matching algorithms.} then virtual views can be synthesized
during video playback using texture and depth maps of the closest
captured camera views (\textit{i.e.}, \textit{anchor views}) via 
depth-image-based rendering (DIBR)~\cite{mark97}.
This ability to construct and observe any virtual view is called
\textit{free viewpoint video}~\cite{tanimoto11}, which enables a 3D visual 
effect known as \textit{motion parallax}~\cite{zhang09}: a viewer's 
detected head movements trigger correspondingly shifted video views on 
his/her 2D display. It is well known that motion parallax is the 
strongest cue in human's perception of depth in a 3D 
scene~\cite{reichelt10}, enhancing the immersive experience.

In a live free viewpoint video streaming scenario, texture and depth
videos from multiple viewpoints in the same 3D scene are real-time
encoded into separate streams at server before delivery to 
interested peers. The clients, organized in a P2P system, can 
choose to look at the recorded 
anchor views or virtual views that are arbitrarily positioned 
between the anchor views. 
Because the distortion of synthesized view tend to be 
larger as virtual view distance to anchor views 
increases~\cite{cheung11tip2}, it is beneficial for a viewer to 
request anchor views that 
tightly ``sandwich'' the virtual viewpoint he wants 
to look at. 
On other hand, given that a group of local peers can share the access
cost of common anchor views, peers have incentive to collaboratively
select and share the same anchor views, even if doing so means that the
anchor views are further away with a distortion penalty in
the synthesized views.  In this paper, we investigate the anchor view
allocation problem for collaborative streaming of live free viewpoint
video under different network settings. To the best of our knowledge,
this is the first piece of work addressing such an issue for
collaborative streaming of free viewpoint video.

As a peer changes his interested view $u$ over time, $u$ may eventually
move outside the viewing range $[v^l,v^r]$ delimited by his two
current anchor views $v^l$ and $v^r$. This necessitates the
system to reallocate new anchor views for the peer.  If
such network reconfiguration costs due to peers' view-switching is
negligible, we first show that the anchor view allocation problem can
be efficiently and optimally solved in polynomial time using dynamic
programming (DP). This is true no matter if the anchor view access
cost from the server to the group of peers is formulated as 
a constraint (\textit{i.e.}, the maximum number of anchor views
allocated to a peer group cannot be larger than a certain number
$B_{\max}$) or as a cost function (\textit{i.e.}, each anchor view
pulled from the source incurs a certain access cost $a$).

On the contrary, if the network reconfiguration cost is
non-negligible due to peers' view-switching, (\textit{e.g.}, in 
the case of large or frequent view-switching by the peers),
the problem of anchor view allocation becomes NP-hard for both 
formulations of anchor view access cost (as a constraint or as a cost 
function). We thus present a locally optimal  
and centralized allocation algorithm inspired by the Lloyd's algorithm
in non-uniform  
scalar quantization~\cite{vq92}. Finally, we propose a distributed 
version of the algorithm with guaranteed 
convergence, where each peer group can independently makes
merge-and-split decisions with a well-defined fairness criteria.  The
results show that our proposed algorithms achieve optimal and
close-to-optimal performance respectively in terms of total cost, and 
substantially outperform a P2P scheme without collaborative 
anchor selection.

The outline of the paper is as follows. We first discuss related 
work in Section~\ref{sec:related}. We then overview the live free
viewpoint video streaming in Section~\ref{sec:system}. We first
formulate the anchor view allocation problem with negligible network 
reconfiguration cost 
and the
corresponding optimal DP algorithm 
in Section~\ref{sec:formulate}. We then 
formulate our problem with reconfiguration cost in 
Section~\ref{sec:formulate2} and show it is NP-hard. We then describe 
locally optimal solutions to the problem in Section~\ref{sec:algo2}.
Finally, we present results and conclusion in Section~\ref{sec:results}
and \ref{sec:conclude}, respectively.

%% file: related.tex
Though much research in multiview video has been focusing on 
compression (\textit{e.g.}, multiview video coding (MVC) 
\cite{flierl07}),
streaming strategies and network optimization for
multiview video is still a relatively 
unexplored and new research topic. \cite{cheung11tip}
discusses an   
\textit{interactive multiview video streaming} (IMVS) video-on-demand
scenario, where only a single requested view per client is needed at 
one time during video playback as the client periodically requests
view-switches. It proposes an 
efficient coding structure where a captured image can be encoded into 
multiple versions, so that the appropriate version can be transmitted 
depending on the currently available content in decoder's buffer, 
in order to reduce server transmission rate. 
Later, \cite{huang12} leverages on the IMVS coding structure 
for content replication, so that suitable 
versions of multiview video segments can be cached in a 
distributed manner across cooperative network servers.

Our current work on anchor view allocation differs from the above work
in that: i) we consider the more general {\em free} viewpoint video,
where, a client can select and synthesize any
intermediate virtual view between two anchor views via DIBR; and ii) we
focus on the {\em live collaborative} streaming scenario, where anchor
views can be shared among peers that are synchronized in time
but not necessarily in view.

There has been a large body of work on peer-to-peer (P2P) streaming,
addressing different aspects of the problem.  For example, 
\cite{magharei07, lu10} study the structure and organization of
streaming overlays, while the work of \cite{vu10, chang11}
discuss the design and deployment of large-scale P2P streaming
systems through measurement on real-world streaming systems.
All the previous works above study single view streaming,
and the results cannot be applied to live free viewpoint video 
streaming, where anchor-view selection  is a critical and
challenging issue.

There has been little work studying multiview streaming over P2P
network. For example, the work of \cite{Ding11} proposes a scheduling 
algorithm that allows peers to frequently compute the scheduling of 
multiview segments. \cite{chen10} studies achieving low 
view-switch delay by organizing viewers with different views together.  
These works essentially treat
multiview video as streaming of multiple single-view videos, and it
is not clear how to extend them
to live free viewpoint streaming where anchor-view selection and its
effect on distortion need to be considered. To the best of our
knowledge, this is the first piece of work on collaborative streaming 
of interactive live free viewpoint video. 

%% file: system.tex
\subsection{Network Model}

We model the free viewpoint video distribution network with two nodes:
$S$ is the server node where live video streams originate,
and $P$ is a single node representing a group of local peers with
close geographical or network distance.\footnote{If the peer group 
is too large, sub-division into smaller groups for independent 
content sharing is also possible.  Our current formulation can be
easily extended to this case.
}
The connection between server $S$ and peer group $P$ may be
modeled as a \textit{hard} constraint; \textit{i.e.}, the number of 
anchor views pulled from $S$ by $P$ cannot exceed $B_{\max}$. 
Alternatively, the connection may be modeled as a \textit{soft}
constraint; \textit{i.e.}, each anchor view pulled by $P$ incurs a 
cost $a$ in the total cost function. The different connection 
constraints are used later in the problem formulation.

\subsection{Free Viewpoint Video Model}
\label{subsec:videoModel}

Let $\mathcal{V} = \{1,2,...,V\}$ be a 
discrete set of \textit{captured viewpoints} for $V$ equally spaced 
cameras in a 1D array as done in \cite{fujii06} and others. Each 
camera captures both a texture (RGB image) and depth map (per-pixel 
physical distances between captured objects in the 3D scene and 
capturing camera) at the same resolution. Texture map from an 
intermediate \textit{virtual viewpoint} between any two cameras 
can be synthesized using  texture and depth maps of the two 
camera views (\textit{anchor views}) 
via a depth-image-based rendering (DIBR) technique like 3D 
warping~\cite{mark97}. Disoccluded pixels in the synthesized view---pixel 
locations that are occluded in the two anchor views---can be filled 
using a depth-based inpainting technique like~\cite{oh09pcs}.

More specifically, denote a virtual viewpoint by $u$ that a peer
currently requests for observation.  We write $u$ as
$u=1+\frac{k}{K}$, $k=\{0, \ldots, (V-1)K\}$, for some large
$K$.\footnote{Though we consider here equally spaced virtual views 
for ease of exposition, our analysis and algorithms can be easily
generalized to uneven virtual view spacing as well.}  In other
words, $u$ belongs to a discrete set of intermediate viewpoints
between (and including) captured views $1$ and $V$, spaced apart by
integer multiples of distance $1/K$ ($u$ approaches a continuum as
$K$ increases).  We consider that a distribution function $q_u$
describes the fraction of peers in the group who currently request
virtual view $u$. Any virtual view $u$ can be synthesized using left
and right anchor views denoted as $v^l$ and $v^r$, respectively, where
$v^l, v^r \in \mathcal{V}$ and $v^l \leq u \leq v^r$. Note that $v^l$
and $v^r$ do not have to be the closest captured views to $u$. The
distortion of the synthesized view varies with the choices of anchor
views.  Let $D_u(v^l, v^r)$ be the distortion function of peers
requesting virtual view $u$, which is synthesized using $v^l$, $v^r$
as anchors.

\subsubsection{Monotonic Distortion model}

A reasonable assumption on distortion is monotonicity with respect to
anchor view distance \cite{cheung11tip2}. It is not guaranteed that 
distortion always decreases with the distance between reference views, but 
this is true in the vast majority of the settings. We hence consider 
a monotonic distortion model in this paper: further-away 
anchor view does not lead to smaller resulting synthesized 
view distortion:
\begin{eqnarray}
D_u(v', v^r_u) & \geq & D_u(v, v^r_u), ~~ \forall v' < v < u \nonumber \\
D_u(v^l_u, v) & \leq & D_u(v^l_u, v'), ~~ \forall u < v < v'
\end{eqnarray}

\subsection{View-switching Model}

To model the view-switching behavior of peers, we consider that a peer
with current desired virtual view $u$ can switch in the next time
instant to any
virtual views $w$'s with probability $P_{u,w}$,
and $\mathbf{P}$ is the view-transition probability matrix. For example,
if a peer stays in the current view 
$u = 1 + k/K$
with probability 
$\Omega$, and switches to any one of the two adjacent views with equal 
probability $(1-\Omega)/2$, we have
the following transition probabilities:
\begin{equation}
P_{1 + k/K, w} = \left\{ \begin{array}{ll}
\Omega & \mbox{if} ~~ w = 1+k/K \\
(1-\Omega)/2 & \mbox{if} ~~ w = 1 + (k \pm 1)/K \\
0 & \mbox{o.w.}
\end{array} \right.
\label{eq:reconfigure}
\end{equation}

%% file: formulate.tex
In this section, we consider the case where the reconfiguration cost
due to peers' anchor view changes is negligible, {\it e.g.}, 
peers tend to switch views infrequently, and hence the
distribution network does not need to be reconfigured often. We 
now formulate the anchor view allocation problem formally as the
\textit{interactive free-viewpoint live streaming} (IFLS) problem.

\subsection{Optimization and System Variables}

We first define the optimization variables, which are the same
for all our formulations of the problem. 
Let $\mathcal{V}' \subseteq \mathcal{V}$ be a \textit{purchased set} of 
captured views selected by the peer group to serve as anchor views to
synthesize virtual views requested. A peer of virtual
view $u$ selects left and right anchor views,  
$v_u^l$ and $v_u^r$ from the purchased set $\mathcal{V}'$ to synthesize 
its desired virtual view $u$.
We consider the following anchor view selection constraint:
\begin{equation}
v^l_u \leq u \leq v^r_u,  ~~~ 
v^l_u, v^r_u \in \mathcal{V}' \subseteq \mathcal{V}, 
~~~ \forall u
\label{eq:anchor}
\end{equation}
In words, Equation~(\ref{eq:anchor}) states that peer of virtual view $u$ 
must select from $\mathcal{V}'$ the left anchor view $v_u^l$ to the left 
of $u$ (\textit{i.e.}, $v_u^l \leq u$) and right anchor view $v_u^r$ to 
the right of $u$ (\textit{i.e.}, $u \leq v_u^r$). The selected anchor views
$v_u^l$, and $ v_u^r$ will induce synthesized distortion $D_u(v^l,
v^r)$, as discussed in Section~\ref{subsec:videoModel}.  These are our
variables to be optimized.

There is an access cost to purchase the set $\mathcal{V}'$ of anchor
views by the peer group $P$.
If there is a hard connection constraint (or cost budget), we have
\begin{equation}
| \mathcal{V}' | \leq B_{\max}
\label{eq:hardCon}
\end{equation}
One may alternatively consider a soft connection constraint, where the 
total access cost $A_{total}$ for the peer group  is proportional to
the number of anchor views purchased, {\it i.e.}, 
$A_{total} =a ~ | \mathcal{V}' |$.
For now, we are only concerned with the access cost of camera
views in the purchased set $\mathcal{V}'$; the question of how the 
cost should be fairly distributed to each peer is deferred to
Section~\ref{sec:fairness}.

 
If the connection is modeled as a hard constraint, the objective of 
the IFLS problem is to select a subset $\mathcal{V}'$ 
and anchor views $v_u^l, v_u^r \in \mathcal{V}'$ for each virtual 
view $u$, so as to minimize the aggregate distortion of all peers of all 
virtual views $u$'s, {\it i.e.}, 
\begin{equation}
\min_{\mathcal{V}' \subseteq \mathcal{V}} ~
\sum_{u} q_u D_u(v_u^l, v_u^r),
\label{eq:IFLS-H}
\end{equation}
subject to Constraints~(\ref{eq:anchor})
and~(\ref{eq:hardCon}).
We label this combinatorial optimization problem as \textit{IFLS-H}.

Alternatively, if the connection is modeled as a soft constraint, the 
objective becomes the combination of total distortion of all peers of 
all virtual views $u$'s \textit{plus} the total access cost, 
\begin{equation}
\min_{\mathcal{V}' \subseteq \mathcal{V}} ~
\sum_{u} q_u D_u(v_u^l, v_u^r) + A_{total}
\label{eq:IFLS-S}
\end{equation}
subject to Constraint~(\ref{eq:anchor}).
We label this problem as \textit{IFLS-S}.

%% file: algorithm.tex
Both IFLS-H and IFLS-S can be solved optimally in polynomial
time via DP.  We show here how IFLS-S is solved;
algorithm for IFLS-H follows similar steps in a straight-forward manner,
and hence is omitted.

Define $\varphi(v^l, u^l, u^r, v^r)$ as the minimum cost for all peers 
interested in virtual views $u \in [u^l, u^r]$, where $v^l$ and 
$v^r$ are the nearest left and right anchor views that have already
been purchased. The optimal solution of IFLS-S can be found by
a call to $\varphi(v_i^l, u_i^l, u_i^r, v_i^r)$, where $u_i^l$ 
and $u_i^r$ are the leftmost and rightmost virtual views requested
by the peer group, and $v_i^l$ and $v_i^r$ are the corresponding
camera views just to the left and right of them, {\it i.e.},
\begin{eqnarray}
v_i^l & = & \left\lfloor u_i^l \right\rfloor, ~~~~~
u_i^l = \arg \min ~\{ u \} ~~~\mbox{s.t.}~~ q_u > 0; \nonumber \\
v_i^r & = & \left\lceil u_i^r \right\rceil, ~~~~~
u_i^r = \arg \max ~\{ u \} ~~~\mbox{s.t.}~~ q_u > 0.
\end{eqnarray}

Given above, $\varphi()$ can be recursively calculated as 

\vspace{-0.1in}
\begin{small}
\begin{eqnarray}
\varphi(v^l, u^l, u^r, v^r)  & = & 
\min \left\{ \sum\limits_{u^l \leq u \leq u^r}q_u d_u(v^l, v^r), 
\min\limits_{v^l < v < v^r} \left[~a~ + \right. \right. \nonumber \\
& &  \left. \left. \varphi(v^l, u^l, u^{v-}, v) ~+~ 
\varphi(v, u^{v+}, u^r, v^r)\right] \right\},
\label{eq:dpdef}
\end{eqnarray}
\end{small}\noindent
where $u^{v-}$  is the virtual view of a peer to the left 
and nearest to new anchor view $v$ ($u^{v-} \leq v$), and $u^{v+}$  is the  
virtual view of a peer to the right and nearest to $v$ $v
< u^{v+}$. The loop
invariant of Equation~(\ref{eq:dpdef}) is $v^l \leq u^l \leq u^r \leq
v^r$.  

In words, Equation~(\ref{eq:dpdef}) states that $\varphi()$ is the
smaller of:
\begin{itemize}
\item[i)] Sum of synthesized distortion of virtual views $u$'s, 
$u^l \leq u \leq u^r$, given that no more anchor views will be purchased
(and hence $v^l$ and $v^r$ are the best anchor views for synthesis of
views $u \in [u^l, u^r]$).
\item[ii)] Cost of one more anchor view $v$, 
$v^l < v < v^r$, which is the access cost $a$ plus the recursive cost
$\varphi()$ using two virtual-view ranges, given by $[u^l, u^{v-}]$ and 
$[u^{v+}, u^r]$, that divide the original range $[u^l, u^r]$. 
\end{itemize}

The complexity of the solution given by Equation~(\ref{eq:dpdef}) can
be analyzed as follows.  Each time Equation~(\ref{eq:dpdef}) is solved
for arguments $v^l, u^l, u^r$, and $ v^r$, they can be stored in entry
$[v^l][u^l][u^r][v^r]$ of a DP table $\Phi$ so that any subsequent
repeated sub-problem can be simply looked up. Each computation of
Equation~(\ref{eq:dpdef}) takes $O(V)$ steps, and the size of the
table is $O(V^4)$.  This results in run-time complexity of
$O(V^5)$.

%% file: formulate2.tex
As the video is played back, a peer may switch his observation
viewpoint from a virtual view $u$ to a new view $u'$, where $u'$ may
fall outside the range $[v_u^l, v_u^r]$ spanned by the anchor views
$v_u^l$ and $v_u^r$. The network hence needs to be reconfigured 
to supply the peer with new anchors.
If the reconfiguration cost is non-negligible, the peer group would tend
to choose anchors $v_u^l$ and $v_u^r$ 
that are further apart, so that the likelihood of the virtual view 
switching outside the range $[v_u^l, v_u^r]$ is low.  In this section,
we formulate the anchor-view allocation problem with reconfiguration
costs, termed \textit{free-viewpoint live streaming with view-switching} 
(FLSV).

\subsection{Reconfiguration Cost}

We define the \textit{reconfiguration cost} $S_u(v_u^l, v_u^r)$ as
the probability that a peer requires new anchor views
during the next $\tau$ view-switches, given the current virtual view $u$
and the anchor views $v_u^l$ and $v_u^r$.  $S_u$ may be computed as
follows. We first define a sub-matrix $\mathbf{P}(v_u^l,v_u^r)$ that
contains only entries $P_{w,z}$'s, where $w, z \in [v_u^l,v_u^r]$, 
defined in Equation~(\ref{eq:reconfigure}).
Note that unlike $\mathbf{P}$, the sum of the entries in a row in
$\mathbf{P}(v^l,v^r)$ does not need to add up to $1$.  We can 
write $S_u$ as a simple sum:
\begin{equation}
S_u(v_u^l, v_u^r) = 1 - \sum_{w} P^{\tau}_{u, w}(v_u^l,v_u^r),
\label{eq:stability}
\end{equation}
where $P^{\tau}_{u,w}(v_u^l,v_u^r)$ is the entry $[u][w]$ in matrix 
$\mathbf{P}^{\tau}(v_u^l,v_u^r)
= \prod_{t=1}^{\tau} \mathbf{P}(v_u^l,v_u^r)$, the $\tau-$step
transition probability. 
In words, Equation~(\ref{eq:stability}) states that the
reconfiguration cost  
$S_u$ is one minus the probability that the peer stays within the range 
$[v_u^l,v_u^r]$ for all $\tau$ view switches.


\subsection{Objective Function}

We first consider the server-peer cost as a hard constraint, and
formulate the {\it FLSV-H} optimization problem.  The objective is to 
select a subset $\mathcal{V}'$ of camera views and to select anchor views
$v_u^l, v_u^r$ for each virtual view $u$
within $\mathcal{V}'$, in order to minimize the 
total distortion of all peers plus a reconfiguration cost weighted by
$\mu$, {\it i.e.},
\begin{equation}
\min~~ \sum_{u} q_u (D_u(v^l, v^r) + \mu S_u(v^l, v^r)),
\label{eq:FLSVobj}
\end{equation}
subject to Constraints~(\ref{eq:anchor}) and~(\ref{eq:hardCon}).

We next consider the connection as a soft constraint.  The objective
then becomes the sum of the distortion, reconfiguration
cost, \textit{plus} total access cost, {\it i.e.},
\begin{equation}
\min_{\mathcal{V}' \subseteq \mathcal{V}} ~
\sum_{u} q_u (D_u(v^l, v^r) + \mu S_u(v^l, v^r)) + A_{total},
\label{eq:FLSV-S}
\end{equation}
subject to Constraint~(\ref{eq:anchor}).
This problem is \textit{FLSV-S}.

\subsection{NP-Hardness Proof}

Both FLSV-H and FLSV-S are NP-hard.  We present the proof of FLSV-H
here; the proof of FLSV-S follows similar argument and is discussed
in the Appendix. 

We show that the well known NP-complete {\em Minimum Cover} (MC)
problem is polynomial-time reducible to a special case of FLSV-H.  In 
MC, a collection $\mathcal{C}$ of subsets of a finite item set
$\mathcal{S}$ is given. The decision problem is: does $\mathcal{C}$
contain a \textit{cover} for $\mathcal{S}$ of size at most $\kappa$, 
{\it i.e.}, a subset $\mathcal{C}^\prime \subseteq \mathcal{C}$ where
$|\mathcal{C}^\prime| \leq \kappa$, such that every item in $\mathcal{S}$
belongs to at least one subset of $C^\prime$?

Consider a special case of FLSV-H where in the optimal solution,
all peers use the leftmost camera view 1 as their left anchor 
view. This is the case if the synthesized distortion for each peer of
view $u$ is a local minimum whenever view 1 is used as left anchor, 
\textit{i.e.}, $D_u(1, v^r_u) \leq D_u(v, v^r_u), \forall v, v^r_u$.
Hence all peers will share view 1 as left anchor view, and need 
to select only right anchor view to minimize the aggregate cost in 
Equation~(\ref{eq:FLSVobj}).

\begin{figure}
	\centering
	\includegraphics[width=2.8in]{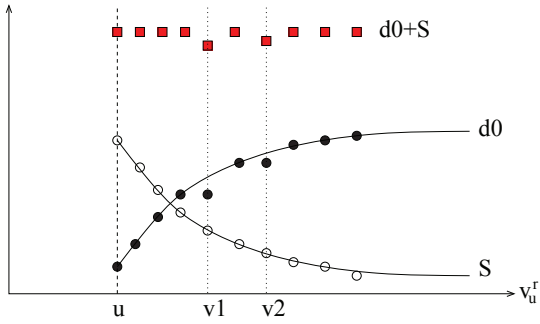}
	\caption{Cost with different right anchor.}
	\label{fig:aggcost}
\end{figure}

We first map items in set $\mathcal{S}$ to consecutive virtual views 
$u$'s (each with $q_u = 1/|\mathcal{S}|$) just to the right of 
leftmost captured view $1$. We map subsets in collection $\mathcal{C}$ 
to captured views $v$'s to the right of the virtual views $u$'s. 
We next construct reconfiguration cost $S_u(1,v^r_u)$ by assuming a 
view-switching probability $\Omega > 0$ in (1) and $\tau=1$, resulting 
in a decreasing $S_u(1,v^r_u)$ as function of $v^r_u$ for all virtual 
views $u$'s, as shown in Figure~\ref{fig:aggcost}.

We first set distortion $D_{u}(1,v_u^r)$ for peers of virtual views $u$'s
such that the aggregate cost is a constant $\alpha$, \textit{i.e.}, 
$D_{u}(1,v_u^r) + S_{u}(1,v_u^r)= \alpha$. Then for each item $s_i$ in 
subset $c_j$, we reset distortion $D_{u}(1,v_u^r)$ (of virtual view 
$u$ corresponding to item $s_i$ and of anchor view $v_u^r$ corresponding
to set $c_j$) to distortion $D_{u}(1,v_u^r-1)$ of anchor view $v_u^r-1$. 
Note that the distortion function remains monotonically non-decreasing.

Figure~\ref{fig:aggcost} shows an example of the aggregate
cost for peer of virtual view $u$, where $d_0$ is the distortion
and $S$ is the reconfiguration cost. Note that $d0+S = \alpha$ except 
for $v_u^r = v_1$ and $v_u^r = v_2$.  
If an optimal solution to FLSV-H with constraint $V_M = \kappa+1$ has 
a total cost less than $ |\mathcal{S}| \alpha$, then the 
selected camera views will correspond to $\mathcal{C}'$ in
$MC$. Hence MC is a special case of FLSV-H.  $\Box$

%% file: algo2.tex
In this section, we present heuristic algorithms to address the anchor view
selection problem with reconfiguration cost.  We first
present a centralized and locally optimal algorithm based on Lloyd's
algorithm~\cite{vq92} in non-uniform scalar quantization. Then we present 
a distributed algorithm with guranteed convergence, followed by the fair
access cost allocation mechanism.

\subsection{Local Optimum with Lloyd's Algorithm}

We present here a low-complexity centralized optimization algorithm
that converges to a locally optimal solution for FLSV.  We first 
observe that for a given subset $\mathcal{V}' \subseteq
\mathcal{V}$ of camera views with a fixed access cost
$A_{total}$, a peer of virtual view $u$ can \textit{independently}
select 
$v_u^l$ and $v_u^r$ from
$\mathcal{V}'$ in order to minimize its own sum of distortion and
reconfiguration cost given by $D_u(v^l, v^r) + \mu S_u(v^l, v^r)$.  This
potentially 
leads to a better global solution. In other words, a solution cannot
be globally optimal if a peer of a view $u$ can lower his own sum of
distortion and reconfiguration cost by choosing a different left or
right anchor views from the same purchased set $\mathcal{V}'$. We
formalize this necessary condition for global optimality
in the following lemma.

\begin{lemma}
If $\mathcal{V}'$, $v_u^l$'s and $v_u^r$'s are a set of optimal
variables, then peer(s) of any virtual view $u$ cannot switch from a
selected left anchor view $v = v_u^l$ to another anchor view
$v' \in \mathcal{V}'$ and lower the overall cost. $\Box$
\end{lemma}

The above Lemma also holds for switching of right anchor view to lower
overall cost.

While the first lemma is concerned with switching of anchor views
within a fixed subset $\mathcal{V}'$ of camera views, we can
similarly construct a second Lemma concerning a selected camera view
$v \in \mathcal{V}'$ being replaced by another camera view
$v' \not\in \mathcal{V}'$.

\begin{lemma}
If $\mathcal{V}'$, $v_u^l$'s and $v_u^r$'s are a set of optimal
variables, then one cannot replace a selected camera view
$v \in \mathcal{V}'$ with an unselected camera view
$v' \not\in \mathcal{V}'$, so that peers of views $u$'s that currently
select camera view $v$ as anchor, i.e. $v_u^l = v$
or $v_u^r = v$, switch to $v'$ as anchor, and lower overall cost. $\Box$
\end{lemma}

These two Lemmas are analogous to the two necessary conditions in 
optimizing non-uniform scalar quantization (SQ). SQ is the problem of 
quantizing a large number of samples in $R^1$ space into $k$ Voronoi 
regions for compact representation, so that only $\lceil \log k \rceil$ 
bits are required to represent a sample with minimal distortion. The first 
necessary optimal condition for SQ is that each sample can freely select 
a Voronoi region to represent itself, one whose centroid has the minimum 
distance to itself (minimum distortion). This is similar to our first 
Lemma.  In the second optimal condition, each Voronoi region can freely 
select a centroid that minimizes the sum of distance to all samples in 
the region. This is similar to our second Lemma.

Due to the similarity of our problem to SQ, we can deploy a
modified version of the famed Lloyd's algorithm to solve our problem.
We call our algorithm the \textit{centralized peer grouping} (CPG) 
algorithm. 

For FLSV-H, we first
pull the leftmost and rightmost camera views from the server, and
then a total number of $B_{max} - 2$ camera views are randomly
pulled in between.  For each peer we calculate the optimal anchor views
(chosen from $B_{max}$ camera views) that minimizes the sum
of its distortion and reconfiguration cost.  Similar to the Lloyd's
algorithm, we iteratively adjust the positions of $B_{max} - 2$
camera views to reduce the total costs of all peers in the group.
In each iteration, we go through each one of $B_{max} - 2$ camera
views, calculate the new total costs if we shift the camera view one
step towards its left and right. If the new total cost is lower than
the original, we substitute the camera view with the one to its left
(or right).  The algorithm stops when the total cost of peers cannot
be further reduced. It is guaranteed to converge since the total cost
only decreases in each iteration.

For FLSV-S, we run the above procedure $V-1$ times with $B_{max} = 2$
to $V$, and then choose the optimal $\mathcal{V}'$ that gives us the minimum
total cost due to distortion, reconfiguration and access.

\subsection{Distributed Heuristic}
\label{sec:dh}
\input{heuristic.tex}

%

%

\subsection{Fair cost allocation within a coalition}
\label{sec:fairness}
\input{faircostallocation.tex}

%% file: heuristic.tex
The centralized algorithm presented above is able to find a nearly
optimal FLSV solution by assigning anchor views to each
peer. The solution is suitable when there is a central controller, and
the network is not large or highly dynamic (with peer arrivals, many
view switchings and departures). In this section, we present a simple,
adaptive and distributed heuristic for collaborative sharing of anchor 
views, or equivalently for constructing the overlay P2P network, which 
scales well to large network with peer churns.
We call this distributed heuristic the 
\textit{distributed peer grouping} (DPG) algorithm. 



\begin{figure}
	\centering
	\includegraphics[width=3in]{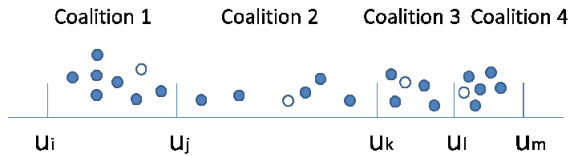}
	\caption{Coalition of peers.}
	\label{fig:coalitioning}
\end{figure}

In a peer group, peers watching the same or adjacent virtual views are
organized into ``coalitions''. Figure~\ref{fig:coalitioning} shows an
example of how the peer coalitions are formed, where $u_i, u_j, ..., u_m$ are virtual views. Peers watching virtual views between $u_i$ and $u_j$ are organized into a coalition, i.e., Coalition 1.
All peers that belong to the same coalition share anchor views and thus access
costs. There is a leader peer (marked in white) in each coalition, which keeps track of the
number of peers watching each virtual view and of the total cost of the
whole coalition.  It periodically exchanges the cost information with
both neighoring coalitions on each side. Two neighboring
coalitions may merge into a new bigger coalition, and a coalition may
also split into two coalitions if the overall cost can be reduced.
We discuss algorithms for peer joins, coalition merge
and split, peer leaves and view switching in the following.

\noindent
\textit{Peer Join:}
When a new peer $i$ arrives, it first contacts a {\em Rendezvous
Point} ({\em RP}) that forwards it to the peer group that $i$
belongs to. This could be done with an IP address lookup.  If there is an
existing coalition $g$ that covers the virtual view peer $i$ requests
in the peer group, {\em RP} connects $i$ with the leader node of the
coalition $C$. The node $i$ joins coalition $C$ and starts to pull anchor
views from other peers in the coalition. The leader peer of $C$
updates the cost and information of the coalition. However, if the 
virtual view requested by peer $i$ is not in the range of any coalition, 
a new coalition will be created, and $i$ becomes the leader of
the coalition. It pulls the anchor views from the streaming server
that minimizes its own costs (distortion and reconfiguration cost).

\noindent
\textit{Coalition merge:}
The coalition structure is adaptive to peer churns, which keeps the
P2P network optimized.  The leader peers of each coalition
periodically exchange information with neighboring leaders.  Let
$L_1$, $L_2$ be the cost for $C1$ and $C2$ respectively, and $L_M$ be
the optimal cost from the result of the CPG algorithm run on $C_1 \cup
C_2$ if $C1$ and $C2$ merge and cooperate.  If $L_M < L_1 + L_2$, the
two coalitions $C_1$ and $C_2$ are merged.  Let $V_M$ be the optimal
set of anchor views returned by the CPG algorithm. Each peer $i$
in the merged coalition adapts to new anchor views $v^{l\ast}_i$
and $v^{r\ast}_i$ that give the minimum cost ($v^{l\ast}_i,
v^{r\ast}_i\in V_m$).  The leader who requested the merge becomes the
new leader of the merged coalition.

\noindent
\textit{Coalition split:}
For a big coalitation $C_M$, the leader
periodically examines whether splitting into two coalitions leads to
lower cost.  Let $u_m$ be a virtual view separating $C_M$ into two
coalitions $C_L$, $C_R$.  For each different $u_m$, the leader runs the CPG
algorithm on both $C_L$ and $C_R$. If the combination of optimal costs is smaller than $L_m$, then $C_M$ is split into $C_L$ and $C_R$, and a
new leader will be randomly selected for the newly created coalition.

\noindent
\textit{Peer leave:}
When a peer $i$ is about to leave, all content sharing between $i$ and
its neighbors is stopped, and the leader node updates the cost of the
coalition. If the leader node leaves, a new leader is randomly chosen.

\noindent
\textit{View switch: }
A peer $i$ could switch the virtual view it currently watches in the
middle of a streaming session. If the new virtual view is still
within the range of the coalition, peer $i$ can still pull anchor views
from other peers and synthesize the new view. There
will be no change of the overlay structure. However, if the new virtual
view goes out of the range of the coalition, the peer will leave the
current coalition and join (or create) a new coalition. It follows the
same process as in the situation where peers join or leave the system.

%% file: faircostallocation.tex
We propose a mechanism to fairly distribute the access costs to each peer
for the DPG algorithm described in section~\ref{sec:dh}.
From the above discussion, cooperation enables peers watching adjacent
views to share the anchor views and thus the access cost. It helps to
reduce the total cost of all users. As peers in P2P networks are
selfish and rational, an important issue in our live free viewpoint video
streaming problem is the fair allocation of the cost among peers in a
coalition, so that our solution does not only minimize the total
cost of the entire P2P network, but also helps each user to lower its own
cost. As such, no user is willing to deviate from the proposed
solution, and the constructed overlay P2P network is stable.

Coalitional game theory provides an ideal tool to provide fair rules
for cost reduction via cooperation in our free-viewpoint live
streaming problem \cite{owen95}. Consider a coalition $C=\{1,
2, \cdots, n\}$ with $n$ peers who watch neighboring views and share
the anchor views and the access cost. Let $S \subseteq C$ be a
subgroup of users in $C$ watching nearby views, where $L(S)$ is the total cost
of peers in $S$ if they decide to cooperate, with L being the cost function defined in (\ref{eq:FLSV-S}). An \emph{allocation}
vector $\mathbf{x}=[x_1, x_2 ,\cdots, x_n]$ divides the total cost
$L(C)$ among its $n$ members, where $x_i$ is the cost (including view
distortion, access cost and reconfiguration cost) assigned to user
$i$.\footnote{Note that from Section~\ref{subsec:videoModel} and
Equation~(\ref{eq:stability}), users' view distortion and
reconfiguration costs are fixed once the set of anchor views is selected, and only their access costs can be adjusted to
achieve fairness among peers in a coalition. In our work, given the
desired allocation $\mathbf{x}$, we adjust users' access costs to
ensure that user $i$'s total cost is $x_i$.} 

Given an allocation $\mathbf{x}$, define the \emph{excess} of a
subgroup $S \subseteq C$ (with respect to $\mathbf{x}$) as
$e(S,\mathbf{x}) = L(S)-\sum_{i\in S} x_i$, which is the extra cost
incurred to $S$ if they deviate from the coalition $C$ and the
allocation $\mathbf{x}$ but form a coalition $S$
themselves. If $e(S,\mathbf{x}) > 0$, the subgroup $S$ has
no incentive to deviate from the coalition $C$. For an allocation, 
if its excesses are all non-negative, then users in C have an incentive
 to stay in $C$, and our goal is to find such stable
coalitions and allocations.

Finding such stable allocations is often difficult, and a well known
fair solution is the \emph{nucleolus} \cite{owen95,faigle01}. The
nucleolus always exists and is unique. It maximizes the excesses in
the non-decreasing order, or equivalently, minimizes peers'
dissatisfaction in the non-increasing order. Moreover, it is one of the stable allocations if they exist.
 The nucleolus is defined as
follows. Given an allocation $\mathbf{x}$, let $\Phi(\mathbf{x})$ be
the vector of all excesses $\{ e(S,\mathbf{x}), \emptyset \neq S \neq
C \}$ sorted in the non-decreasing order. The nucleolus
$\mathbf{\eta}$ is the unique allocation that lexicographically
maximizes $\Phi$ over all allocations, that is,
$\Phi(\mathbf{\eta}) \succ_{lex} \Phi(\mathbf{x}), \forall \mathbf{x}\neq \mathbf{\eta}$.\footnote{A
vector $\mathbf{a}$ is said to be lexicographically larger than vector
$\mathbf{b}$ ($\mathbf{a} \succ_{lex} \mathbf{b}$) if in the first
component that they differ, that component of $\mathbf{a}$ is larger
than that of $\mathbf{b}$.}

To compute the nucleolus, we follow the above definition
and solve a sequence of linear programs as follows \cite{faigle01}. We
first solve the following problem
\begin{eqnarray}
    (LP_1) && \max \ \epsilon \cr
           && \sum_{i \in C} x_i = L(C), \cr
           && \sum_{i \in S} x_i \leq L(S) - \epsilon, \ \forall S \neq \emptyset, S \neq C,
\end{eqnarray}
which adds constraints on the allocation vectors $\mathbf{x}$ to maximize the smallest excess. Let $\epsilon_1$ be the optimal solution of ($LP_1$), which is the maximal smallest excess, and let $S_1$ be the collection of all subgroups whose excesses are equal to $\epsilon_1$. We then solve
\begin{eqnarray}
    (LP_2) && \max \ \epsilon \cr
           && \sum_{i \in C} x_i = L(C), \cr
           && \sum_{i \in S} x_i = L(S) - \epsilon_1, \ \forall S \in S_1, \cr
           && \sum_{i \in S} x_i \leq L(S) - \epsilon, \ \text{otherwise},
\end{eqnarray}
which maximizes the second smallest excess. We continue this way until there is only one allocation $\mathbf{x}$ that satisfies all the constraints in the optimal solution, and that allocation is the nucleolus.

In DPG, we apply the above procedure to compute the nucleolus for each coalition found by the algorithm.

%% file: results.tex
In this section we present illustrative simulation results.
In simulations, we assume the distortion function $D_u$ has the following form:
\begin{equation}
D_u(v^l, v^r) = \gamma e^{\alpha_u(v^r - v^l)} 
\left( e^{\beta_u * \min(u-v^l, v^r-u)} - 1 \right).
\label{eq:distort}
\end{equation}
Note that if virtual view $u$ is actually one of the anchor views, then the distortion $D_u$ is zero. The rate at which the distortion increases with the distance between anchor views, depends on the parameters $\alpha_u$ and $\beta_u$. 

Unless otherwise stated, we use the following baseline parameters in our simulation: number of captured views: 21, number of virtual views: 200, number of peers: 10000, $\omega = 0.4, \tau = 6, a = 5$.
We assume that the distribution of peers watching each virtual view follows a  normal distribution. 
We have also run our simulations on different peer distributions.
The results of those simulations are qualitatively the same as what is presented here, and hence are not shown for the sake of brevity.


\subsection{Results for Negligible Reconfiguration Cost}
We compare the DP-based optimal solution with a simple P2P approach for solving the IFLS problem.
In the latter simple P2P approach, peers independently choose the anchor views that minimize their own distortion. The access costs of each anchor view are shared by all users that request it.
There is no collaboration on anchor selections among peers.

\begin{figure*}[!ht]
    \begin{minipage}[c]{0.33333\textwidth}
       \centering
		\includegraphics[width= 1\textwidth]{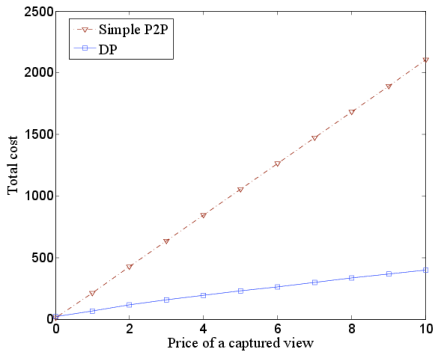}
	\caption{Total cost versus price of captured views.} 
	\label{fig:dp}
    \end{minipage}
    \begin{minipage}[c]{0.33333\textwidth}
       \centering
		\includegraphics[width= 1\textwidth]{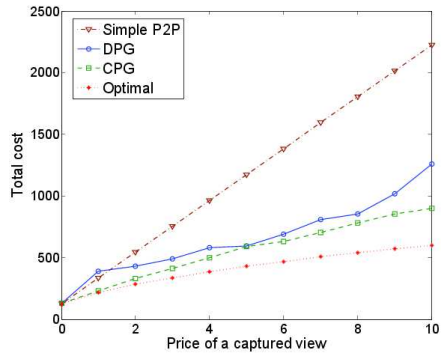}
	\caption{Total cost versus price of captured views.} 
	\label{fig:cost}
    \end{minipage}
		\begin{minipage}[c]{0.33333\textwidth}
       \centering
		\includegraphics[width= 1\textwidth]{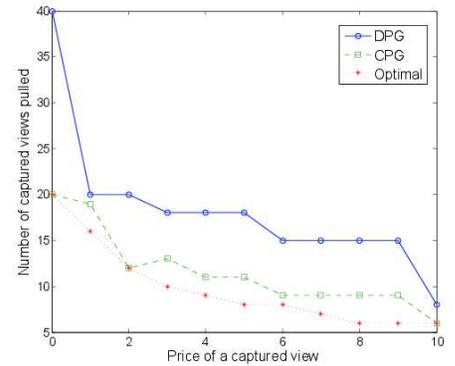}
	\caption{Number of captured views pulled.} 
	\label{fig:anchors}
    \end{minipage}
\end{figure*}

Figure~\ref{fig:dp} shows the total cost (distortion plus access costs) for
the peers as a function of the price of camera views. It is shown that our CPG algorithm gives much better results than the simple P2P approach,
especially when the price is high. This is because, in
the DP algorithm, the peers can collaboratively select and share the same
anchor views to reduce the access cost, with a small price in distortion
penalty. Therefore, fewer captured views are pulled from the server, and
the total cost is minimized. 

\subsection{Results for Non-Negligible Reconfiguration Cost}

We carried out simulation to evaluate the performance of our proposed
CPG and DPG algorithms with the optimal solution (\textit{Optimal}),
and the simple P2P approach.  The optimal solution is obtained through
exhaustive search.  The simple P2P approach is similar to the one we
used in IFLS except that peers choose anchor views to minimize their
own total cost.




Figure~\ref{fig:cost} shows the total cost of all peers versus the price of a captured view. 
It is shown in the figure that the total cost increases with the price of a camera view. This is because a higher view price leads to a higher access cost, and peers tend to share the same anchor views with others so they can share 
the cost of common anchor views from the streaming server. This, in turn, 
increases other cost components, \textit{i.e.}, distortion and 
reconfiguration costs. From the figure, we see that CPG performs very 
close to the global optimal solution. The anchor views can successfully 
adapt to good positions to minimize the total costs of all peers. DPG 
is also very efficient in reducing the total cost, especially when the 
price of a captured view is high. DPG does not outperform {\em simple P2P} 
when the view price is low due to the lack of global information.


Figure~\ref{fig:anchors} shows the total number of views pulled from the 
streaming server as a function of access cost of an anchor view. The number 
drops with the increase in the price of access cost. When requesting a 
captured view from the streaming server becomes expensive, in order to 
reduce their access costs, peers tend to seek more cooperation by using 
the same anchor views and sharing the access cost. Therefore, the total 
number of camera views pulled from the streaming server becomes smaller.
In DPG, the total number of views pulled could be higher than the total 
number of camera views since peers only share the access costs within 
the same coalition, and a captured view could be pulled multiple times 
by peers from different coalitions.

\begin{figure*}[!ht]
    \begin{minipage}[c]{0.33333\textwidth}
       \centering
		\includegraphics[width= 1\textwidth]{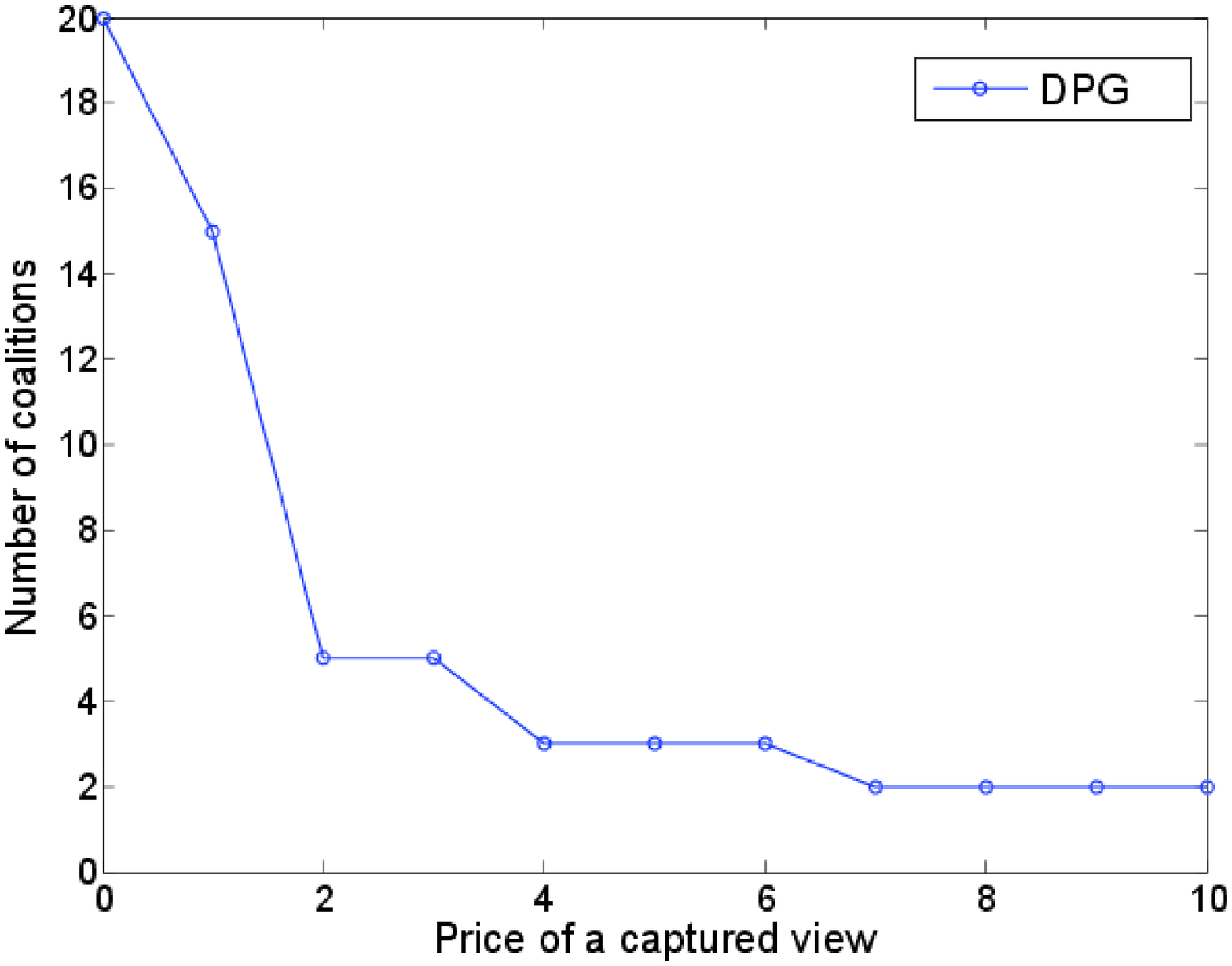}
	\caption{Number of coalitions formed.} 
	\label{fig:coalitions}
    \end{minipage}
	\begin{minipage}[c]{0.33333\textwidth}
       \centering
		\includegraphics[width= 1\textwidth]{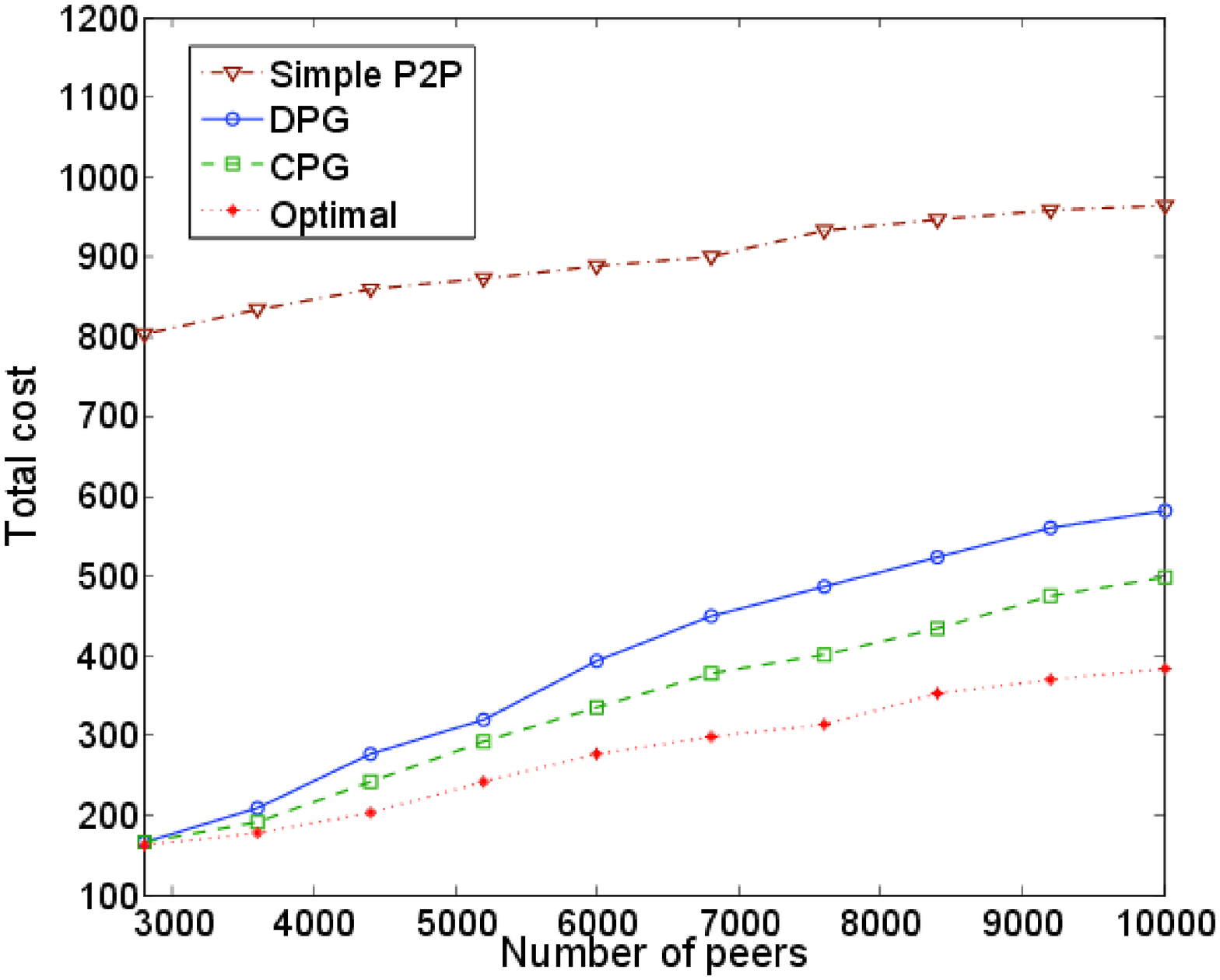}
	\caption{Total cost versus number of peers.} 
	\label{fig:views}
    \end{minipage}
    \begin{minipage}[c]{0.33333\textwidth}
       \centering
		\includegraphics[width= 1\textwidth]{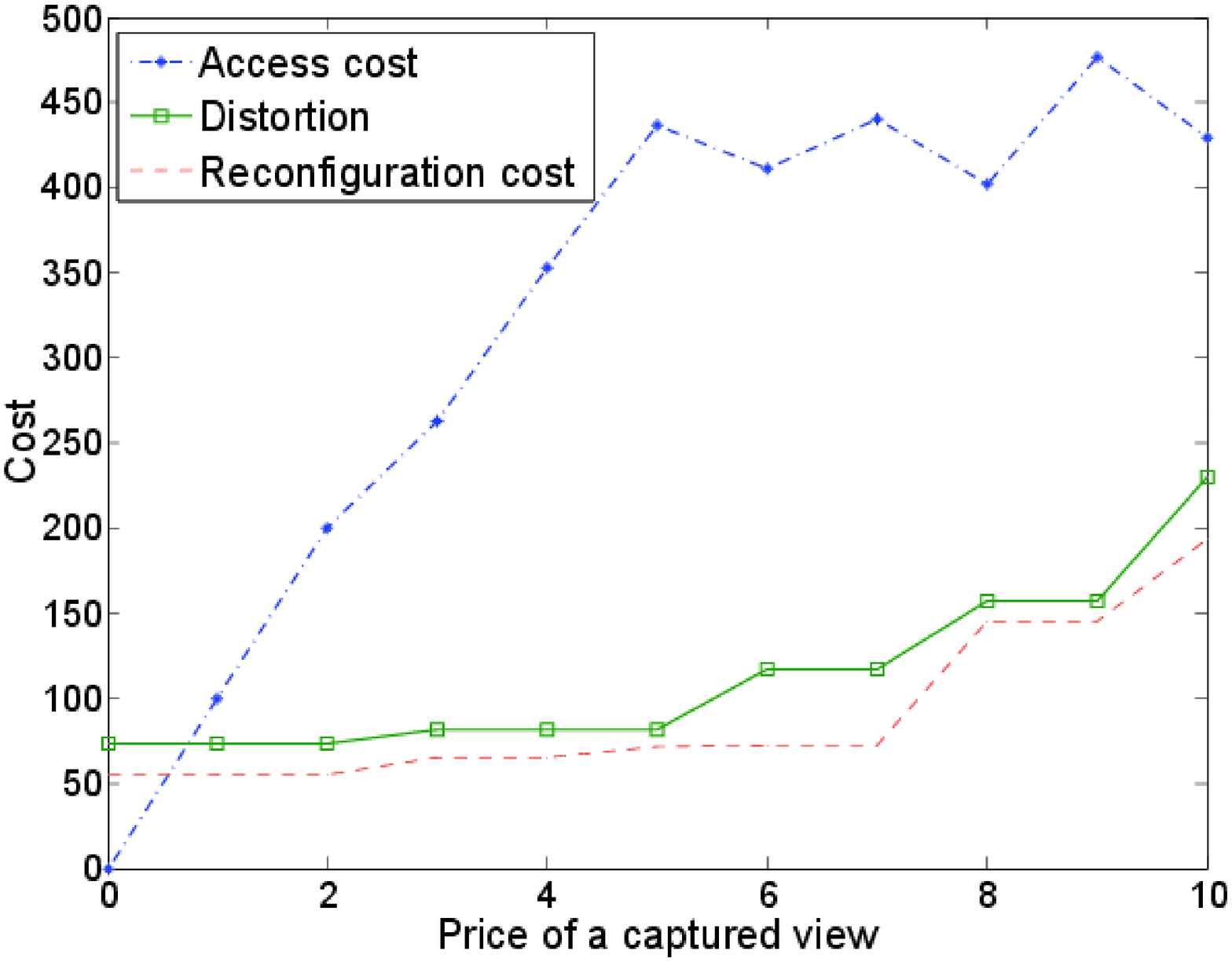}
	\caption{Cost component versus anchor price.} 
	\label{fig:components}
    \end{minipage}

\end{figure*}

Figure~\ref{fig:coalitions} shows the number of coalitions formed by {\em Heuristics} algorithm. The number of coalitions drops with the price of a captured view. When the anchor views are expensive, neighboring coalitions are more likely to merge into a bigger one so that the access costs could be shared by more peers. The {\em Heuristics} can efficiently re-arrange the topology to minimize the total cost when the view prices changes.

Figure~\ref{fig:views} shows the total cost of all peers versus peer population. The total cost increases with the number of peers. Simple P2P performs the worst. It has very high total cost even when the number of peers is low. This is due to the lack of collaboration in peer anchor selections. DPG and CPG achieve close-to-optimal performance. When there are fewer peers in the system, they 
tend to use same anchor views to reduce access cost, with a penalty in 
other cost components. When the peer population increases, each peer can 
choose better anchor views that leads to a lower distortion and 
reconfiguraiton cost, since there will be more neighbors to share the 
access costs.

Figure~\ref{fig:components} shows the cost components of CPG algorithm. 
With the increase of view price, access cost becomes the major component 
of the total cost. Distortion and reconfiguration costs also increase 
because peers compromise to suboptimal anchor views (in terms of 
distortion and reconfiguration) so that their access costs can be 
shared with a larger crowd. The cost components of DPG are qualitatively 
the same as CPG, and hence are not shown for brevity.






%% file: conclude.tex
In this paper we study the design and optimization of interactive P2P 
streaming of live free viewpoint video.  
In free viewpoint live streaming, peers could select different virtual 
viewpoints, which are synthesized using texture and depth videos of the 
anchor views captured by multiple cameras. 
The access cost of common anchor views are collectively shared by peers 
with a price of higher distortion.
We formulate two problems, IFLS with negligible reconfiguration cost, and 
FLSV with none-negligible reconfiguration cost.
Then we provide a DP-based optimal solution for IFLS, and heuristic 
algorithms for FLSV.
The simulation results show that our proposed algorithms achieve 
respective optimal and close-to-optimal performance in terms of total 
cost, and substantially outperform a P2P scheme without collaborative 
anchor selection.

%% file: append1.tex
We prove that FLSV-S is also NP-hard, by reducing the
NP-complete MC problem to a special case of FLSV-S.  Following
similar construction in the proof for FLSV-H, we first map items in set
$\mathcal{S}$ to virtual views $u$'s (each with $q_u =
1/|\mathcal{S}|$) to the right of leftmost captured view $1$, and
map subsets in collection $\mathcal{C}$ to captured views $v$'s to the
right of the virtual views.  Consider again the case where the 
optimal solution has all peers sharing view $1$ as their left anchor.  

We construct reconfiguration cost $S_u(1,v^r_u)$ as done in the 
FLSV-H proof. Next, we identify the smallest $S_u(1,v)$ for all $u$'s 
and $v$'s for which $u$ and $v$ correspond to an item and a subset in 
original MC problem, respectively. Let $\delta = S_u(1,v-1) - S_u(1,v)$. 
We then construct $D_u(1,v)$ to be $1-S_u(1,v)-\delta$ if
the subset corresponding to $v$ contains the item corresponding to $u$,
and $1-S_u(1,v)$ otherwise. That means that a virtual view covered
by a camera view $v$ will have a decrease of $\delta$ in distortion.
Note that by definition of $\delta$, $D_u(1,v)$ is monotonically 
non-decreasing. 
Finally, we define the access cost $a = \delta/(|\mathcal{C}|+1)$, which
means that purchasing all the captured views $v$'s is cheaper than
paying for $\delta$ for a virtual view $u$ uncovered by a captured 
view $v$. 

We now claim that, if the optimal solution to FLSV-S has access cost
smaller than $\kappa \delta / (|\mathcal{C}|+1)$, then the corresponding MC
decision problem is positive, and vice versa.  The reason is the
following. Under the above construction, FLSV-S can always find a
solution that covers all virtual views $u$'s (items in MC) with
camera views $v$'s. If the minimum cost solution requires $\kappa$ or
fewer captured views, then the corresponding subsets will cover all
items in $\mathcal{C}$ in MC.